# Wastewater Pipe Defect Rating Model for Pipe Maintenance Using Natural Language Processing

Sai Nethra Betgeri[1], Shashank Reddy Vadyala[1], John C. Mattews[2], Hongfang Lu[3]

1. *Graduate Assistant, Department of Computational Analysis and Modeling, Louisiana Tech University, Ruston, LA, United States*
2. *Director, Trenchless Technology Center, Ruston, LA, United States*
3. *Associate Professor, Southeast University, Nanjing, Jiangsu, China*

**Abstract:**

Closed-circuit video (CCTV) inspection has been the most popular technique for visually evaluating the interior status of pipelines in recent decades. Certified inspectors prepare the pipe repair document based on the CCTV inspection. The traditional manual method of assessing structural wastewater conditions from pipe repair documents takes a long time and is prone to human mistakes. The automatic identification of necessary texts has received little attention. Computer Vision based Machine Learning models failed to estimate structural damage because they are not entirely understood and have difficulty providing high data needs. Hence, they have problems providing physically consistent findings due to their high data needs. Currently, a very small curated annotated image and video data set with well-defined, precisely labeled categories to test Computer Vision based Machine Learning models. This study provides a valuable method to determine the pipe defect rating of the pipe repair documents by developing an automated framework using Natural Language Processing (NLP) on very small curated annotated image , video data and more text data. The text used in this study is broken into grammatical units using NLP technologies. The next step in the analysis entails using words to find the frequency of pipe defect and then classify into respective defect rating for pipe maintenance. The proposed model achieved 95.0% accuracy,94.9% recall, 95% specificity, 95.9% precision score, and 95.7% F1 score, showing the potential of the proposed model to be used in large-scale pipe repair documents for accurate and efficient pipeline failure detection to improve the quality of the pipeline.

Keywords: Defect detection, Wastewater pipe inspection, Natural language processing (NLP), Text recognition

## 1. Introduction:

The underground pipeline system forms a significant part of the infrastructure because it includes thousands of miles in the United States. Sanitary wastewater collects wastewater from public and private users as part of wastewater infrastructure systems[1, 2]. About 500,000 miles of private wastewater laterals and 800,000 miles of municipal wastewater lines. [3]. By 2032, 56 million people are expected to use centralized treatment plants[4-6]. Water supply and wastewater water pipelines are essential for society's survival, and their security and efficiency are critical for human health and economic growth[7-15]. Using risk-based asset management, the most critical assets to take the most efficient course of action are identified by prioritizing the highest risk of failure by considering all the parameters.

Using the traditional manual method, the number of failures received by wastewater management can increase rapidly, making pipe failure handling imperative because the inspectors manually produce them by checking through CCTV films, manually recognizing and classifying such failures through pipe repair paperwork, and extracting the information connected to those pipe failures is difficult. As a result, the manual extraction procedure has a high potential for human mistakes, time consumption, and information loss. This issue can be resolved by substituting autonomous computational extraction for these manual procedures. The difficulty of computationally extracting information from free-text narratives is addressed by the field of study known as information extraction (IE), a subfield of natural language processing (NLP).

NLP can be an efficient way of automatically extracting information from large-scale pipe repair documents. NLP is a computer-assisted approach for facilitating the processing of human (natural) language. NLP may be a research area developing techniques accustomed to analyzing and extracting valuable data from text and speech in natural languages. Several NLP applications include information extraction, language translation, and opinion mining [16-20]. Text classification using NLP has been used for many problems within pipeline construction: To classify construction documentation based on priority[21, 22]. To support field inspection and data extraction of the inspection[23]. Information from work hazard analyses is processed using an ontology-based text categorization approach [24]. NLP methods are often divided into two main categories: (1) Rule-based and (2) Machine learning (ML) [25]. Systems that rely only on hand-coded syntactic rules are known as rule-based systems[26, 27]. As a result, their performance is underwhelming. Languages and linguistic grammar are unimportant in the Machine Learning-based approach[28] because patterns are often quickly learned from unclear training examples because they outperform the state-of-the-art model like $K$-NN.
Nevertheless, NLP is a valuable technique for extracting and processing information from natural language into a more organized format for study. Natural Language Toolkit (NLTK) systems may be automated to parse textual content and search for keywords and phrases to extract data using predefined computer algorithms. The following is how key phrase extraction may be expressed as a sequence labeling task. Predict a sequence of labels, one label for each word in the input, where each label key phrase word) or non-KP, given an input sequence x, where each x represents the input vector of the keyphrase word). The task formulation for sequence labeling considers the correlations between nearby labels. It enables the joint decoding of the optimal sequence of labels for the input sequence rather than decoding each label individually.

Inspired by advancements in natural language processing, some researchers have more recently applied recurrent neural networks (RNNs) to wastewater pipe assessment documents for extraction and classification [29-33]. Recurrent neural networks (RNNs) include Long Short-Term Memory Networks (LSTMs) that solve the problem of RNNs' gradient disappearing. Additional memory cells in LSTMs are used to store memories from long-distance phrases. Because LSTMs may store information from past sequence inputs in the current input state, they have proven a natural option for data applications such as speech recognition, language modeling, and trial option [34]. An LSTM has a hidden layer, an input layer, and an output layer[35]. The hidden state in a forward LSTM network only saves information from the past. With the regular LSTM, input flows either in backward or forward directions. In bidirectional, input flows in two directions, creating a Bi-LSTM different from the regular LSTM. A bi-directional LSTM network with a forward hidden layer and a backward hidden layer to capture information flow in both directions is utilized [36-38]. The first model learns the sequence of the input provided, and the second model learns the reverse of that sequence. Data in the model is unstructured data to extract information from both sides at the entity level. The nodes in the hidden layer are linked, which is how long-distance information is kept in the matrix weights.

A comparison is performed between an LSTM and Bi-LSTM in pipe defect rating applications. LSTM and Bi-LSTM used many-to-many configurations for flaw detection and localization on simulated ultrasonic A-scans of holes and cracks. In their case, each exhibited perfect performance in the outputs of an LSTM in a many-to-many configuration as input to a dense decision-making layer for defect extraction, and assigning defect ratings on wastewater pipe assessment documents has historically proven to be challenging. It is unknown how these models generate particular decisions of defect classification and rating assignment because it is tough to interpret these data-driven models and how the rating is assigned to each defect characteristic. In addition, these methods are trained on small, curated data, and their generalization ability on unseen data is often limited [39, 40].

## 2. Objective

This paper proposed an ontology-based framework to improve efficiency and support decision-making regarding extraction and assign defect rating by automating the text classification by considering a complete set of defect Lexicon. The proposed pipe defect rating model uses the deep representation of entities using a knowledge base to reduce human efforts for labeled data creation and feature engineering. To illustrate the effectiveness of the proposed model, empirical experiments are conducted on a real dataset from the Department of Engineering & Environmental Services in Shreveport, Louisiana.

## 3. Methods and Materials

### 3.1 Data Set and Data Preprocessing:

A total of 3100 pipe repair documents were extracted from the Dept. of Engineering & Environmental Services' approved database by removing records with insufficient and missing information for further analysis. There was no complete information about the defect location, so 130 documents were reoved, and finally had 2970 pipe repair documents. Table 1 shows the information included in the pipe repair document.

Table 1: Description of pipe inspection document

| Section | Description |
|---|---|
| Pipe Characteristics | Information about the physical pipe properties (Ex: Diameter, Depth, Length) |
| Emergency Repair | Information about the Emergency Repair (Ex: Immediate Leakage fixes) |
| Smoke Testing Assessment | Information about any smoke Observed from pipes (Ex: Medium smoke observed emanating from Cleanout) |
| Defects | Information about the pipes using CCTV Cameras (Ex: Multiple Defects) |
| Composite Assessment | Information about the Composite Material around the pipe |
| Criticality Assessment | Information about the risk value of the pipe (Ex: Medium) |
| Capacity | Information about the pipe capacity |
| Summary | Information about total Major and Minor Defects |

A PACP incorporated Comprehensive rating protocol was established to provide a standardized way of documenting features and assign defect ratings during the inspection to schedule maintenance. Comprehensive and PACP allows wastewater professionals to classify, evaluate and manage inspection data. The most used defect scale is the Comprehensive and PACP Protocol manual listed in Table 2.

Table 2: PACP incorporated Comprehensive rating protocol

| Defect Rating | Description |
|---|---|
| Defect rating 1 | Reassess in ten years |
| Defect rating 2 | Rehabilitate or replace in six to ten years |
| Defect rating 3 | Rehabilitate or replace in three to five years. |
| Defect rating 4 | Rehabilitate or replace in zero to two years. |
| Defect rating 5 | Rehabilitate or replace immediately. |

Next, Data preprocessing for the text from pipe repair documents is performed. Data preprocessing is crucial when dealing with text data because text data is unstructured data and the interpretations of the documents by different inspectors. Next, all special characters in the pipe repair documents (e.g., commas in a list) are removed. Misprint in the dataset included typographical errors (e.g., "Leaks" instead of "Laeks"), and NLTK handles the correction of the spelling errors. Then, boundary detection using Stanford parser's sentence detection is performed since it is a reasonably accurate tool in NLP[3]. Pipe repair documents contain many negation statements (no leaks, no defects...etc.). To identify such terms, entities, or sentences, the Negex function is used [41].

### 3.2 Annotation:

Next, Standard documents are manually annotated for entities and pipe defect ratings. Manual annotation of pipe repair documents provided a gold standard to benchmark the proposed model for pipe defect rating. Two annotators manually mark the list of lexical units and assign a defect rating using the Comprehensive and PACP Protocol to each record. The meaning of sentences and defect ratings can be interpreted differently by different ways inspectors. Annotating final size 2970 pipe repair documents at the entity and record levels and calculating the kappa coefficient for each document and entity. If there is any disagreement between the experts, documents must be annotated again, which is a time taking process. So, a random selection of 500 records is considered a golden standard, annotating each document at the entity and document level. A statistical measure of Cohen's kappa coefficient is used to find the agreement between the two-annotator in Inter Annotator Agreement (IAA)[42]. Inter Annotator Agreement for our entities is shown in Table 3. Kappa coefficients are considered for evaluating agreement coefficients between the expert's opinions compared for each entity level and record level annotation. So, randomly 500 pipe repair documents were selected for the proposed model. 80 percent of the pipe repair documents were used as training data and 20 percent as test data to evaluate the proposed framework for defect rating, as shown in Figure 1.

Table 3: Annotator agreement table

| Entities and defect rating | IAA |
| --- | --- |
| Defect | 0.78 |
| Size defects | 0.82 |
| Locations of defect | 0.84 |
| Defect rating 1 | 0.75 |
| Defect rating 2 | 0.79 |
| Defect rating 3 | 0.84 |
| Defect rating 4 | 0.83 |
| Defect rating 5 | 0.92 |

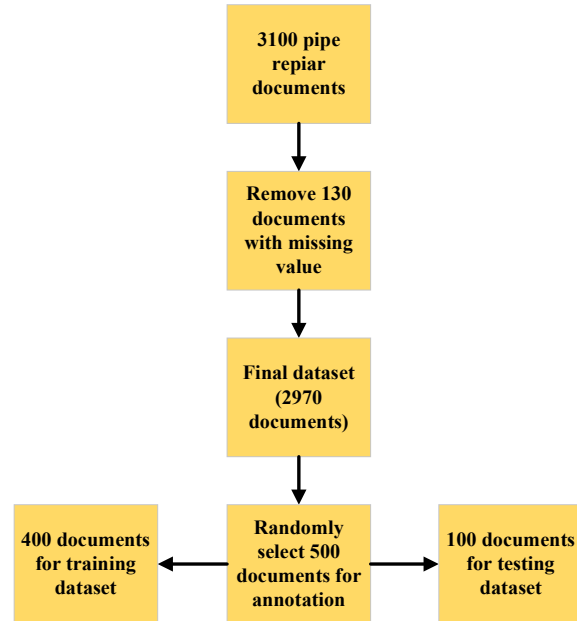
Figure 1. Training and testing data selection.

### 3.3 Lexicon generation

Knowledge bases for pipe defects are built by collecting information from documents and structured sources. The creation of a general list of defect attributes is referred to as a defect lexicon or seed words. Secondly, a list of defects was hand-picked for inspectors from traditional sources. After manually selecting defect attributes, each attribute was expanded with the appropriate verb, noun, and adjective where possible, e.g., leak, leaking, leaked. Then, a knowledge-based method for identifying the new defect attributes is used. Knowledge-based methods exploit available lexicographical resources such as WordNet or HowNet. A lexicon was developed by searching WordNet for a term's synonyms and antonyms [43]. According to [44], the closer two words means fewer iterations are needed to identify the synonymous connection between those two words. The relationship between terms in a knowledge base was employed in both investigations. As illustrated in Figure.2, these systems' basic technique is to use seed sets of pipe fault terms and their orientations to expand this collection of defect characteristics by searching for synonyms and antonyms in a knowledge base.

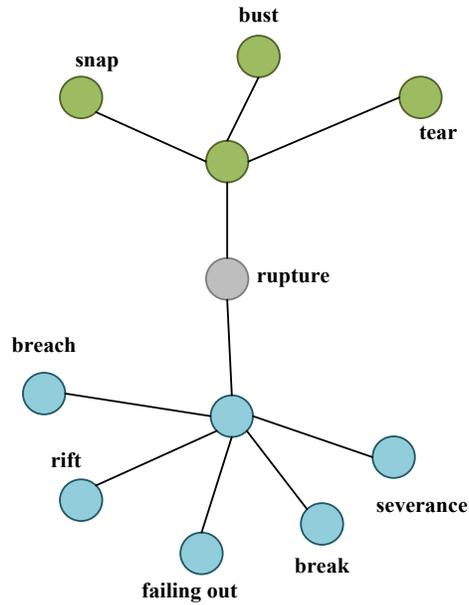

Figure 2. Sample knowledge base construction for defect term "Rupture"

There were a lot of synonyms that needed to be more on-topic and unconnected. After manual tracking of individual seed keywords, a problem is discovered. Their synonyms revealed ambiguous synonyms such as homonyms and antagonyms and incorrect or fuzzily interpreted synonyms. By making different blacklists for each synonym, including redundant, misunderstood, and fuzzily understood terms, the problem is fixed. Unnecessary processing of undesirable terms is avoided in this way. The lexicon knowledge file of pipe defect ontology is shown in Table 4.

Table 4: Description of Pipe defect Ontology

| Location | Frequency of defects | Defects |
|---|---|---|
| Mid-point, Upstream, Downstream, Depth category, Pipe length at longitudinal, Spiral, Circumferential, | Rarely, Several, Frequently, Often, Moderate, Very rarely | Fractures, Sags, Smoke, Leaks, Water level, Sags, Deposits, Joint offset, Deposits attached Encrustation, Deposits settled, compacted, Infill runner, Intruding sealing hanging, Intruding sealing ring loss/poorly fitting, Tap factory defective, Corrosion, Pitting, Gap, Hole, Stain, Rough spot, Foible, Rupture, No defect |

**3.4 Entity Extraction:**

Entity Extraction aims to identify entities mentioned in the text and classify them into predefined entity types, as shown in Table.5. Manual rules are created to fix the problem of dealing with unstructured pipe repair documents from multiple inspectors. Sentences and specific items from the text employed in defect rating calculations are to be identified; for example, "Leak" should be recognized as a defect. Entity extraction graphical representation is shown in Figure.3.

Table 5. Description of the entities

| Entities | Description |
|---|---|
| Defect | Keywords (e.g., Leakage, Rupture, etc.) |
| Location of defect | Keywords (e.g., junction, end, etc.) or distance from the end of the pipe |
| Frequency of defects | Keywords (e.g, Rarely, Frequently etc.) |

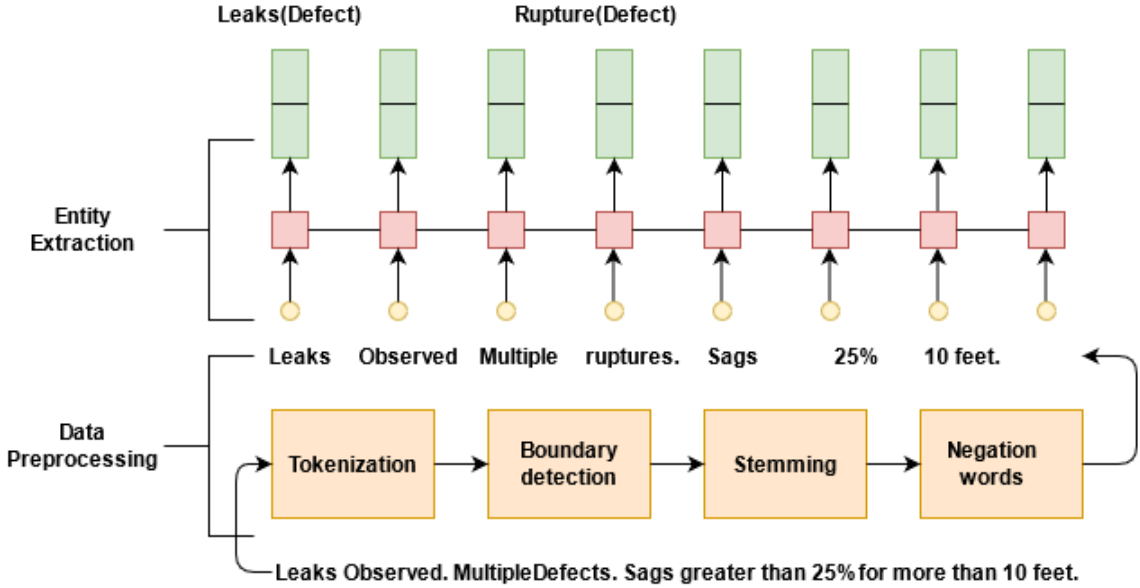

Figure 3. Graphical representation of the overview of entity extraction

Lastly, a Bi-LSTM neural network is implemented. The output vector from both forward and backward sequences is adjoined to obtain the final entity representation vector using the lexicon file generated in section 3.1. The Bi-LSTM model is composed of two LSTM networks and is capable of reading input reviews in both directions, forward and backward. The forward LSTM processes information from left to right and its hidden s,tate and it is shon  s $\vec{h}_t = LSTM(x_t, \vec{h}_{t-1})$ and the backward LSTM processes information by reading from right to left and its hidden state can be expressed as $\overleftarrow{h}_t = LSTM(x_t, \overleftarrow{h}_{t-1})$. Finally, the output of Bi-LSTM can be summarized by concatenating the forward and backward states as $h_t = [\vec{h}_t, \overleftarrow{h}_t]$ and the data frame for each sentence consisting of information such as (defects, size of defect, location of defect, and frequency of defects) are created, which will be used for defect rating calculation. The structure of a Bi-LSTM network is shown in Figure 4. The parameter settings for the entity extraction model are shown in Table 6.

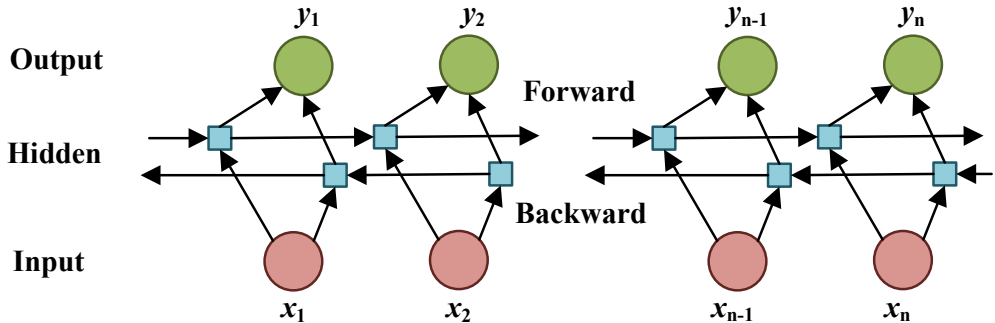

Figure.4. Layers in a Bi-LSTM neural network

Table 6. Parameter setting of the entity extraction model.

| Parameter | Value |
|---|---|
| Word vector embedding size | 200 |
| Dictionary feature vector embedding size | 100 |
| # Hidden neurons for each hidden layer | 300 |
| Batch size | 100 |
| Tag Indices | 4 |
| Learning Rate | 0.005 |
| Number of epochs | 10 |
| Optimizer | Adam optimizer |

## 3.5 Defect Rating:

The proposed defect rating calculation employs three aspects of within pipe repair document defect term frequency ($w_{frequencies}$), the importance of a defect term ($w_{defects}$), and the location of the defect ($w_{location}$) within the pipe, which are developed using term frequency algorithm [45].

The location of the defect plays an important role in defect rating. For locations, the weights are assigned based on the scores mentioned in Pipeline Assessment - NASSCO [32, 40, 46]. Table 7 shows the randomly assigned weights based on the pipe defect location. When no location is found $w_{location}$= 1.0 because it could be entered as multiple locations.  Table 8 shows the randomly assigned weights based on the frequency of defect occurance. Similarly, the defects factor is randomly chosen based on the pipe failure. Table 9 shows the random weights assignment based on defect lexicon units found. Table 10 shows the defect ratings assigned based on $w_{frequencies}, w_{location}, w_{defect}$.

Table 7: Weights assignment based on location of the defect.

| Location | Assigned weight |
|---|---|
| One location | $w_{location}$= 0.9 |
| Multiple locations | $w_{location}$= 1.0 |
| No location | $w_{location}$= 1.0 |

Table 8: Weights assignment based on frequency of the defects outcome

| Frequency outcome | Assigned weight |
|---|---|
| Very rarely or none | $w_0 = 0.1$ |
| Rarely | $w_1 = 0.25$ |
| Moderate | $w_2 = 0.50$ |
| Moderate to Frequently | $w_3 = 0.75$ |
| Frequently, More or very Frequently/ Several/ Oftenly | $w_4 = 0.99$ |

Table 9: Weights assignment based on lexicon units found

| Lexicon or Defect found | Assigned weight |
|---|---|
| No lexicon unit | $w_{defect} = 0.5$ |
| One lexicon unit | $w_{defect} = 0.8$ |
| Multiple lexicon unit | $w_{defect} = 1.0$ |

Table 10: Defect Rating assignment based on $w_{frequencies}, w_{location}, w_{defect}$

| Defect Rating | $w_{frequencies}, w_{location}, w_{defect}$ |
|---|---|
| Defect rating 1 | $w_{frequencies} = 0.1, w_{location} = 0.9$ or $1.0$, $w_{defect} = 0.5$ |
| Defect rating 2 | $w_{frequencies} = 0.25, w_{location} = 0.9$ or $1.0$, $w_{defect} = 0.8$ or $1.0$ |
| Defect rating 3 | $w_{frequencies} = 0.5, w_{location} = 0.9$ or $1.0$, $w_{defect} = 0.8$ or $1.0$ |
| Defect rating 4 | $w_{frequencies} = 0.75, w_{location} = 0.9$ or $1.0$, $w_{defect} = 0.8$ or $1.0$ |
| Defect rating 5 | $w_{frequencies} = 0.99, w_{location} = 0.9$ or $1.0$, $w_{defect} = 0.8$ or $1.0$ |

Defect Rating Score example, 05CCD pipe inspection document contains the following information, as shown in Box.1.

> Very Frequently, there is a leakage in pipe at 10 feet away from pipe installation

Box 1: Sample pipe repair document

It consists of only one sentence, and the location factor is 0.9 because it has only one location. The $w_{frequencies}$ term weights 0.99 for $w_4$, which matches frequently. The defect term (leakage) matches with the lexicon unit and has a 1.0 (seed term) weight, so the $w_{defect}$ is 0.8
So, for the example shown the rating assigned for Pipe 05CCD is 5, which means it needs an immediate replacement or rehabilitation.

Figure.5 illustrates the entire schema of the proposed framework.

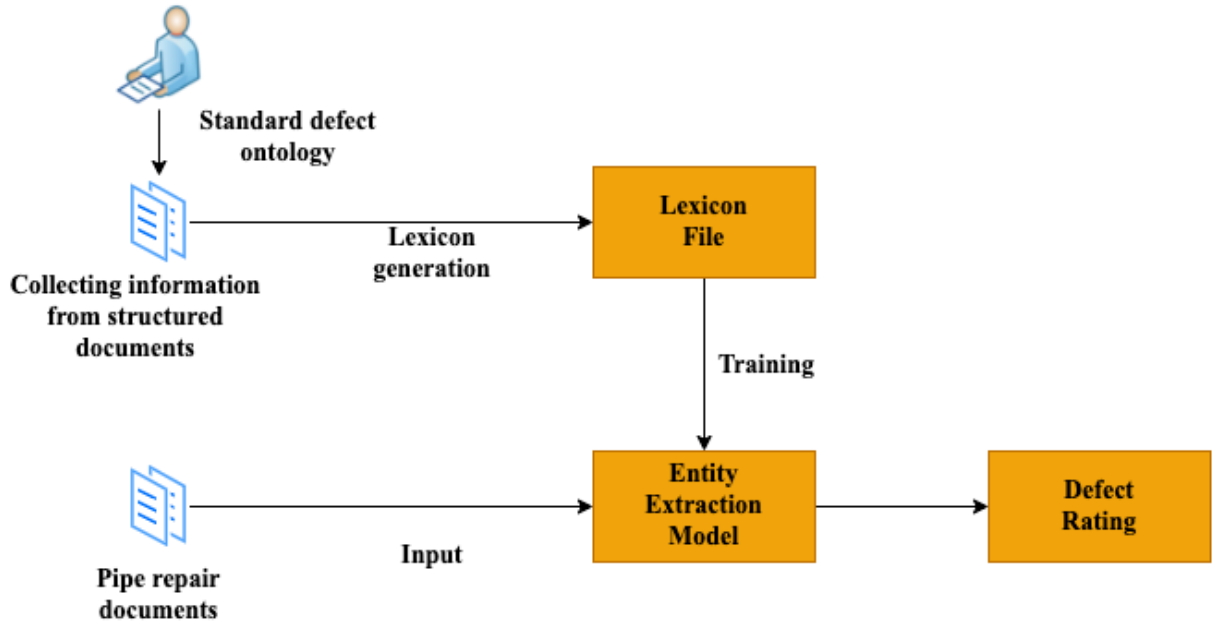

Figure 5. A framework for the proposed defect rating

## 4. Results and Discussions:

The proposed framework defect rating model was trained on the training subset. Finally, the trained models were evaluated on the test subset. To test the defect rating model, precision, recall, accuracy, specificity, and F1 score presented in Eq. 4-8 were utilized [47].

$$Accuracy = \frac{\text{True positive } + \text{ True Negative}}{\text{True positive } + \text{True Negative } + \text{ False Positive } + \text{ False Negative}} \quad \text{Eq.4}$$

$$Precision = \frac{\text{True positive } + \text{ True Negative}}{\text{True positive } + \text{ False Positive}} \quad \text{Eq.5}$$

$$Recall = \frac{\text{True positive}}{\text{True positive } + \text{ False Negative}} \quad \text{Eq.6}$$

$$Specificity = \frac{\text{True Negative}}{\text{True positive } + \text{ False Negative}} \quad \text{Eq.7}$$

$$F1\ Score = 2 \times \frac{\text{Recall } \times \text{ Precision}}{\text{Precision } + \text{ Recall}} \quad \text{Eq.8}$$

To accurately assess the quality of pipes for scheduling maintenance, it is essential to extract detailed information from pipe repair documents using entity extraction. The entity extraction model identifies pipe repair attribute entities from the pipe repair documents. The entity extraction model's overall average accuracy for all entities is 95.1%. The F1 scores of defects, defect size, location of the defect, and frequency of defect terms/entities are 93.6%, 92.4%, 93.6%, and 95.2%, as shown in Table 11. The entity extraction calculates the defect scores and assigns the defect rating.

Table 11. Results of entity extraction model units (%)

| Entity tags | Accuracy | Recall | Specificity | Precision | F1 |
|---|---|---|---|---|---|
| Defects | 92.0 | 94.1 | 92.3 | 95.0 | 93.6 |
| Location of defect | 95.0 | 91.1 | 96.9 | 97.5 | 93.6 |
| Frequency of defects | 96.2 | 95.6 | 93.0 | 94.2 | 95.2 |

The defect statements' representation varied from inspection to inspection. Pipes, for example, can be represented in a variety of ways, such as a 'pipeline,' a 'sewer Line,' or a 'waterline.' The entity extraction model adapted well to different inspector reporting styles and languages. Findings show that accuracy, precision, F1-score, and recall have all improved significantly. The entity extraction accuracy of the Bi-LSTM entity model was consistently higher. Experiments revealed that using sentence boundary detection, entities, and a domain dictionary in the tasks improved accuracy considerably, demonstrating the use of merging techniques and domain dictionaries in entity extraction tasks.

A defect rating must be correct for it to be considered accurate. Precision, recall, accuracy, and F1 score were calculated to evaluate the models' performance. The proposed defect rating model achieved higher accuracy by 97.0%, 92.0%, 93.0%, 95.0%, and 98.0% for the defect rating 1, defect rating 2, defect rating 3, defect rating 4, and defect rating 5, respectively, as shown in Table 9. After a deep analysis of the results, a conclusion was made that misclassified records have complicated sentences, which are very hard for the model to understand and classify, For example. multiple is used for both defects and locations. Table 12 shows the results of the defect rating assignment.

Table 12: Results for defect rating model units (%)

| Defect Rating Score | Accuracy | Recall | Specificity | Precision | F1 |
|---|---|---|---|---|---|
| Defect rating 1 | 97.0 | 97.5 | 97.0 | 97.0 | 97.5 |
| Defect rating 2 | 92.0 | 93.0 | 92.5 | 94.0 | 93.0 |
| Defect rating 3 | 93.0 | 94.0 | 95.5 | 95.5 | 95.5 |
| Defect rating 4 | 95.0 | 93.0 | 92.5 | 95.0 | 94.5 |
| Defect rating 5 | 98.0 | 97.0 | 97.5 | 98.0 | 98.0 |

The results indicate that defect rating can be accurately calculated with the help of the entity extraction model. However, the defect rating model struggled to accurately calculate defect rating pipe repair documents due to fewer defect attributes in pipe repair documents, high disagreement between inspectors who annotated pipe repair documents, and not mentioning the frequency of the defect in a few pipe repair documents.

Waste water pipe maintenance is a critical issue faced by many utilities in the United States. PACP protocol, PACP incorporated Comprehensive rating protocol was developed a few years back to resolve the issue by the utilities. Both developed methods follow manual inspection. The defect occurrence frequency is a vital determinant in determining the severity of the pipe failure. The variety of defects essentially invalidates the assumptions of text mining approaches such as decision trees because a single defect could be divided into multiple categories.

Frequency as an indicator of severity helps to favor the automation process such as tf-idf[48, 49]. tf-idf determines the uniqueness of a text through frequently or rarely. tf-idf would detect the pipe failure but not the severity of the pipe failure and tf-idf. In our approach, the words often, frequently, rarely, etc capture the frequency of the defect. Negation and Location play a vital role not considering them could create data noise.

A disadvantage of the Machine Learning method approach is focusing on correlations, which means results are based on statistics. An inspector may have difficulty understanding the correlations; however, a grammatical approach used in this work can be more understandable.

## 5. Strength and Limitations:

This study is groundbreaking in several ways. This is the first study to look at the validity of NLP for numerous defect entities while investigating pipe repair papers for defecting ratings. Secondly, NLP research in pipe repair documents concentrates exclusively on a single defect condition, such as leaks. With the help of lexicon generation using WordNet, a completed set of defect Lexicon is created and used in defect rating. There are multiple drawbacks also to the proposed model. Firstly, To address accurate defect detection and recognition of wastewater defects that deals with pipe assessment document data, an approach for detecting and recognizing defects, including pipe geographical location, is essential and is not considered in the proposed model. Secondly, the present level of wastewater system management differs from city to city. The algorithm must be trained to improve by collecting more pipe-related lexicon terms from different cities and must be applied to different cities data.

## 6. Conclusions and Future Work:

Since the health state of the wastewater pipe is assessed and recorded as the basis for decision-making in this process, wastewater pipe assessment is the foundation for creating an effective maintenance plan. The proposed model is a reliable and accurate approach to detecting pipe repair documents and assigning defect ratings. The proposed model is suitable for actual pipe repair documents —in terms of performance, inference speed, and simplicity in output interpretation, which can accurately characterize a particular location and defect. This model is tailored to solve the challenges in the wastewater infrastructure. The lexicon generation step is the core for data integration to the proposed model, where the defect ontology entities and semantic rules are developed for representing different types of information related to the specific location, which helps the user to customize the defect lexicon according to location. However, more study is needed to assess the applicability and validity of NLP approaches for trenchless or building projects. In the future, there is a chance to improve the proposed algorithm by collecting more pipe-related lexicon terms from data from different cities and adding pipe-specific geographical locations. Secondly, a framework that consists of data extractions of CCTV videos using Deep

Learning Algorithms and feeds the data for assessing the Suitability of Trenchless Technologies by the Decision Support System for installing, replacing, or rehabilitating each pipe and reducing the methods of costs and evaluating all the 70 technologies and helps in selecting the appropriate techniques would be developed.